\documentclass[12pt]{iopart}
\usepackage[dvips]{graphicx}
\usepackage{ifthen,array}
\usepackage[comma,square,sort&compress]{natbib}
\usepackage{amsfonts}
\usepackage{epsfig}
\usepackage{rotating}

\def\rp{$R_p \hspace{-1em}/\;\:$ }

\def\vb#1{\vbox to #1 pt{}}

\def\half{{\textstyle{1 \over 2}}} 
 
\def\quarter{{\textstyle{1 \over 4}}} 
 
\def\eighth{{\textstyle{1 \over 8}}}

\DeclareMathAlphabet{\mathsc}{OT1}{cmr}{m}{sc}

\def\10{$SO(10)$}
\def\21{SU(2) $\otimes$ U(1) }

\def\lsim{\raise0.3ex\hbox{$\;<$\kern-0.75em\raise-1.1ex\hbox{$\sim\;$}}}
\def\gsim{\raise0.3ex\hbox{$\;>$\kern-0.75em\raise-1.1ex\hbox{$\sim\;$}}}

\def\vb#1{\vbox to #1 pt{}}

\newcommand{\Sol}  {\mathsc{sol}}
\newcommand{\Atm}  {\mathsc{atm}}
\newcommand{\Dma}  {\Delta m^2_\Atm}

\def\21{SU(2) $\otimes$ U(1) }

\newcommand{\Dms}  {\Delta m^2_\Sol}
\newcommand{\flux}[2][]{\ensuremath{\ifthenelse{\equal{#1}{}}{}{^{#1}\!}\mathit{#2}}}

\newcommand{\AddrAHEP}{%
  AHEP Group, Instituto de F\'{\i}sica Corpuscular --
  C.S.I.C./Universitat de Val{\`e}ncia \\
  Edificio Institutos de Paterna, Apt 22085, E--46071 Valencia, Spain}

\begin{document}


\title{ Supersymmetric Origin of Neutrino Mass  }

\author{M.~Hirsch} 
\address{\AddrAHEP} 
\author{J.~W.~F.~Valle} 
\address{\AddrAHEP} 


\begin{abstract}
    Supersymmetry with breaking of R-parity provides an attractive way
  to generate neutrino masses and lepton mixing angles in accordance
  to present neutrino data. We review the main theoretical features of
  the bilinear R-parity breaking (BRpV) model, and stress that it is
  the simplest extension of the minimal supersymmetric standard model
  (MSSM) which includes lepton number violation. We describe how it
  leads to a successful phenomenological model with hierarchical
  neutrino masses.
  In contrast to seesaw models, the BRpV model can be probed at future
  collider experiments, like the Large Hadron Collider or the Next
  Linear Collider, since the decay pattern of the lightest
  supersymmetric particle provides a direct connection with the lepton
  mixing angles determined by neutrino experiments.
\end{abstract}  
\maketitle

\newpage

\setcounter{page}{1} 

\section{Introduction}

A combination of solar, atmospheric, reactor and accelerator neutrino
experiments~\cite{fukuda:1998mi,ahmad:2002jz,eguchi:2002dm} have now
firmly established the existence of neutrino masses and therefore the
incompleteness of the standard model of electroweak interactions.  The
determination of neutrino oscillation parameters presented in
Ref.~\cite{Maltoni:foc} uses the most recent data and state-of-the-art
solar and atmospheric neutrino fluxes. For previous reviews and
references
see~\cite{pakvasa:2003zv,Barger:2003qi,gonzalez-garcia:2002dz,Fogli:2003pz}.
We have now learned that the atmospheric oscillations involving
$\nu_{\mu} \leftrightarrow \nu_{\tau}$ are characterized by a nearly
maximal mixing, while the solar neutrino mixing angle is large, but
significantly non-maximal.  With the recent standard solar model
fluxes there is a unique range for the solar mass splitting $\Dms$,
determined from the data to be about 30 times smaller than the
atmospheric mass splitting $\Dma$.

The discovery of neutrino mass constitutes the only solid hint we
currently have of physics beyond the standard model. There are
theoretical arguments based on the stability of the gauge hierarchy
which suggest the existence of physics at the TeV scale.
Supersymmetry~\cite{nilles:1984ge,haber:1985rc} provides an answer to
both these issues which fits well with unification and string theory
ideas~\cite{Witten:2002ei}.

Prompted by these data there has been a rush of theoretical and
phenomenological papers on models of neutrino masses and mixings.  The
most popular idea is to ascribe neutrino masses to physics at a large
mass scale in order to implement some variant of the see-saw mechanism
\cite{gell-mann:1980vs,yanagida:1979,schechter:1980gr,mohapatra:1981yp}.
Broken R-parity supersymmetry provides a theoretically interesting and
phenomenologically viable alternative to the origin of neutrino mass
and
mixing~\cite{aulakh:1982yn,Ross:1985yg,ellis:1985gi,abada:2001zh,Grossman:1999hc,bednyakov:1998cx,Joshipura:2002fc}.
Here we focus on the case of supersymmetry with bilinear R-parity
breaking~\cite{diaz:1998xc}.  This is the simplest of all R parity
violating models. It also provides the simplest extension of the
MSSM~\cite{diaz:1998xc} to include the violation of lepton number, as
well as a calculable framework for neutrino masses and mixing angles
in agreement with the experimental data
\cite{Diaz:2003as,chun:2002vp,hirsch:2000ef,romao:1999up}. In this
model the atmospheric neutrino mass scale is generated at the
tree-level, through the mixing of the three neutrinos with the
neutralinos, in an effective `low-scale'' variant of the seesaw
mechanism \cite{Ross:1985yg}. In contrast, the solar mass and mixings
are generated radiatively~\cite{hirsch:2000ef}.  BRpV can be
considered either as a minimal extension of the MSSM
\cite{deCampos:1995av,Banks:1995by,deCarlos:1996du,akeroyd:1998iq}
(with no new particles) valid up to some very high unification energy
scale, or as the effective description of a more fundamental theory in
which the breaking of R-parity occurs in a spontaneous way by
minimizing the scalar
potential~\cite{Masiero:1990uj,romao:1992vu,romao:1997xf}.

This short review is mainly devoted to the generation of neutrino
masses and lepton mixing, both the tree-level atmospheric neutrino
mass scale as well as a description of the main features of the full
one-loop calculation of the neutrino-neutralino mass matrix and its
various analytic approximations which, in some cases, can be rather
simple.  For definiteness we will stick to the case of explicit BRpV
only.  

However, in contrast to the seesaw mechanism, in broken R-parity
supersymmetry neutrino masses are generated at the electro-weak
scale~\cite{Diaz:2003as,hirsch:2000ef,romao:1999up}. Such low-scale
schemes for neutrino masses have the advantage of being testable also
outside the realm of neutrino experiments. Although neutrino
properties can not be predicted from first principles, their fit to
the data allows for unambiguous tests of the theory at accelerator
experiments~\cite{Mukhopadhyaya:1998xj,Allanach:1999bf,romao:1999up,Choi:1999tq,bartl:2000yh,porod:2000hv,restrepo:2001me,hirsch:2002ys,Bartl:2003uq,Hirsch:2003fe}.
Indeed, the measured lepton mixing angles lead to well defined
predictions of the decay properties of the lightest supersymmetric
particle (LSP).  This is a very general and robust feature of these
theories, which holds irrespective of the nature of the LSP. Here we
will illustrate possible phenomenological scenarios by discussing some
examples of measurements of decay properties of different LSP
candidates.

This paper is organized as follows.  In Sec.~\ref{sec:formalism} we
introduce the main features of the model, discuss the soft
supersymmetry breaking terms, as well as the relevant fermion
mass matrices and the main features of the corresponding diagonalizing
matrices.  In Sec.~\ref{sec:tree-level-mass} we discuss the generation
of the atmospheric neutrino mass scale at the tree--level, while in
Sec.~\ref{sec:one-loop-induced} we analyse the main features of the
one--loop--induced solar neutrino mass scale, including a discussion
of the relevant Feynman graph topologies. We also give simplified approximation
formula for the solar mixing angle. We then turn briefly to collider 
phenomenology and how the model under discussion could be tested in 
LSP decays in Sec. \ref{sec:test-neutr-prop}
before we conclude and summarize our results in Sec. \ref{sec:disc-concl}.

\section{ Formalism}
\label{sec:formalism}

In this section we introduce the main features of the model and the
relevant mass matrices. The superpotential of the model and the 
soft SUSY breaking terms are given, approximate solutions to the 
tadpole equations discussed.

\subsection{The Superpotential and the Soft Breaking Terms}
\label{TheModel}

The minimal BRpV model we are working with is characterized by the
presence of three extra bilinear terms in the superpotential analogous
to the $\mu$ term present in the MSSM. Using the conventions of
ref.~\cite{akeroyd:1998iq} it may be given as
\begin{equation}  
W=\varepsilon_{ab}\left[ 
 h_U^{ij}\widehat Q_i^a\widehat U_j\widehat H_u^b 
+h_D^{ij}\widehat Q_i^b\widehat D_j\widehat H_d^a 
+h_E^{ij}\widehat L_i^b\widehat R_j\widehat H_d^a 
-\mu\widehat H_d^a\widehat H_u^b 
+\epsilon_i\widehat L_i^a\widehat H_u^b\right] 
\label{eq:Wsuppot} 
\end{equation} 
where the first three terms are the usual MSSM Yukawa terms, $\mu$ is
the Higgsino mass term of the MSSM, and $\epsilon_i$ are the three new
terms which violate lepton number in addition to R--Parity. The
couplings $h_U$, $h_D$ and $h_E$ are $3\times 3$ Yukawa matrices and
$\mu$ and $\epsilon_i$ are parameters with units of mass.  The
smallness of the bilinear term $\epsilon_i$ in eq.~(\ref{eq:Wsuppot})
may arise from a suitable symmetry.  In fact any solution to the $\mu$
problem~\cite{Giudice:1988yz} potentially explains also the
``$\epsilon_i$-problem''~\cite{nilles:1997ij}. A common origin
for the $\epsilon_i$ terms that account for the neutrino oscillation
data, and the $\mu$ term responsible for electroweak symmetry breaking
can be ascribed to a horizontal family symmetry of the type suggested
in Ref.~\cite{Mira:2000gg}.

The smallness of $\epsilon_i$ could also arise dynamically
in models with spontaneous breaking of R
parity~\cite{Masiero:1990uj,romao:1992vu,romao:1997xf}, where it is
given as the product of a Yukawa coupling times a singlet sneutrino
vacuum expectation value.


Supersymmetry breaking is parameterized with a set of soft supersymmetry 
breaking terms. In the MSSM these are given by
\begin{eqnarray} 
{\cal L}_{soft}^{MSSM}&=& 
M_Q^{ij2}\widetilde Q^{a*}_i\widetilde Q^a_j+M_U^{ij2} 
\widetilde U_i\widetilde U^*_j+M_D^{ij2}\widetilde D_i 
\widetilde D^*_j+M_L^{ij2}\widetilde L^{a*}_i\widetilde L^a_j+ 
M_R^{ij2}\widetilde R_i\widetilde R^*_j \nonumber\\ 
&&\!\!\!\!+m_{H_d}^2 H^{a*}_d H^a_d+m_{H_u}^2 H^{a*}_u H^a_u- 
\left[\half M_s\lambda_s\lambda_s+\half M\lambda\lambda 
+\half M'\lambda'\lambda'+h.c.\right]\label{eq:Vsoft} \\ 
&&\!\!\!\!+\varepsilon_{ab}\left[ 
A_U^{ij}\widetilde Q_i^a\widetilde U_j H_u^b 
+A_D^{ij}\widetilde Q_i^b\widetilde D_j H_d^a 
+A_E^{ij}\widetilde L_i^b\widetilde R_j H_d^a 
-B\mu H_d^a H_u^b\right] 
\,.\nonumber 
\end{eqnarray} 
In addition to the MSSM soft SUSY breaking terms in ${\cal
  L}_{soft}^{MSSM}$ the BRpV model contains the following extra terms
\begin{equation}
V_{soft}^{BRpV} = - B_i\epsilon_i\varepsilon_{ab}\widetilde 
L_i^a H_u^b\,,
\label{softBRpV}
\end{equation}
where the $B_i$ have units of mass. In what follows, we neglect
intergenerational mixing in the soft terms in eq.~(\ref{eq:Vsoft}).



The electroweak symmetry is broken when the two Higgs doublets $H_d$
and $H_u$, and the neutral component of the slepton doublets
$\widetilde L^1_i$ acquire non--zero vacuum expectation values (vevs).
These are calculated via the minimization of the effective potential
or, in the diagrammatic method, via the tadpole equations.  The full
scalar potential at tree level is
\begin{equation} 
V_{total}^0  = \sum_i \left| { \partial W \over \partial z_i} \right|^2 
        + V_D + V_{soft}^{MSSM} + V_{soft}^{BRpV}
\label{V}
\end{equation} 
where $z_i$ is any one of the scalar fields in the superpotential in
eq.~(\ref{eq:Wsuppot}), $V_D$ are the $D$-terms, and $V_{soft}^{BRpV}$
is given in eq.~(\ref{softBRpV}).  

The tree level scalar potential contains the following linear terms 
\begin{equation}
V_{linear}^0=t_d^0\sigma^0_d+t_u^0\sigma^0_u+t_1^0\tilde\nu^R_1
+t_2^0\tilde\nu^R_2+t_3^0\tilde\nu^R_3\,,
\label{eq:Vlinear}
\end{equation}
where the different $t^0$ are the tadpoles at tree level. They are given by
\begin{eqnarray}
t_d^0&=&\Big(m_{H_d}^2+\mu^2\Big)v_d+v_dD-\mu\Big(Bv_u+v_i\epsilon_i\Big)
\nonumber\\
t_u^0&=&-B\mu v_d+\Big(m_{H_u}^2+\mu^2\Big)v_u-v_uD+v_iB_i\epsilon_i
+v_u\epsilon^2
\nonumber\\
t_1^0&=&v_1D+\epsilon_1\Big(-\mu v_d+v_uB_1+v_i\epsilon_i\Big)+\half\Big(
v_iM^2_{Li1}+M^2_{L 1i}v_i\Big)
\label{eq:tadpoles}\\
t_2^0&=&v_2D+\epsilon_2\Big(-\mu v_d+v_uB_2+v_i\epsilon_i\Big)+\half\Big(
v_iM^2_{Li2}+M^2_{L2i}v_i\Big)
\nonumber\\
t_3^0&=&v_3D+\epsilon_3\Big(-\mu v_d+v_uB_3+v_i\epsilon_i\Big)
+\half\Big(v_iM^2_{Li3}+M^2_{L3i}v_i\Big)
\nonumber
\end{eqnarray}
where we have introduced the notation
\begin{equation} 
H_d={{H^0_d}\choose{H^-_d}}\,,\qquad 
H_u={{H^+_u}\choose{H^0_u}}\,,\qquad
\widetilde L_i={{\tilde L^0_i}\choose{\tilde\ell^-_i}}\,,
\label{eq:shiftdoub} 
\end{equation} 
and shifted the neutral fields with non--zero vevs as
\begin{equation}
H_d^0\equiv{1\over{\sqrt{2}}}[\sigma^0_d+v_d+i\varphi^0_d]\,,\quad
H_u^0\equiv{1\over{\sqrt{2}}}[\sigma^0_u+v_u+i\varphi^0_u]\,,\quad
\tilde{L}_i^0\equiv{1\over{\sqrt{2}}}[\tilde\nu^R_i+v_i+i\tilde\nu^I_i]\,.
\label{shiftedfields}
\end{equation}

The five vacuum expectation values can be expressed in spherical
coordinates as
\begin{eqnarray} 
v_d&=&v\sin\theta_1\sin\theta_2\sin\theta_3\cos\beta\nonumber\\
v_u&=&v\sin\theta_1\sin\theta_2\sin\theta_3\sin\beta\nonumber\\ 
v_3&=&v\sin\theta_1\sin\theta_2\cos\theta_3\label{eq:vevs}\\
v_2&=&v\sin\theta_1\cos\theta_2\nonumber\\
v_1&=&v\cos\theta_1\nonumber
\end{eqnarray} 
which preserves the MSSM definition $\tan\beta=v_u/v_d$ with the $W$
boson mass given as $m_W^2=\quarter g^2 ( v_d^2 + v_u^2 + v_1^2+
v_2^2+ v_3^2)$.  We have also defined
$D=\eighth(g^2+g'^2)(v_1^2+v_2^2+v_3^2+v_d^2-v_u^2)$ and
$\epsilon^2=\epsilon_1^2+\epsilon_2^2+\epsilon_3^2$. A repeated index
$i$ in eq.~(\ref{eq:tadpoles}) implies summation over $i=1,2,3$. The
five tree level tadpoles $t_{\alpha}^0$ are equal to zero at the
minimum of the tree level potential, and from there one can determine
the tree level vacuum expectation values.

\subsection{Radiative Breaking of the Electroweak Symmetry}
\label{ewbreaking}

A reliable description of electroweak symmetry breaking and Higgs
boson physics in supersymmetry requires the inclusion of radiative
corrections. In the BRpV model the full scalar potential at one--loop
level, called effective potential, is
\begin{equation}
V_{total}  = V_{total}^0 + V_{RC}
\end{equation}
where $V_{total}^0$ is given in Eq.~(\ref{V}) and $V_{RC}$ include the
quantum corrections. Following Refs.~\cite{Diaz:2003as,hirsch:2000ef}
we use the diagrammatic method, incorporating the radiative
corrections through the one--loop corrected tadpole equations. The one
loop tadpoles are
\begin{equation}
t_{\alpha}=t^0_{\alpha} -\delta t^{\overline{DR}}_{\alpha}
+T_{\alpha}(Q)=t^0_{\alpha} +\widetilde T^{\overline{DR}} _{\alpha}(Q)
\label{tadpoles}
\end{equation}
where $\alpha=d,u,1,2,3$ and $\widetilde T^{\overline{DR}} _{\alpha}(Q)
\equiv-\delta t^{\overline{MS}}_{\alpha}+T_{\alpha}(Q)$ are the finite one 
loop tadpoles. At the minimum of the potential we have $t_{\alpha}=0$, and 
the vevs calculated from these equations are the renormalized vevs.

Neglecting intergenerational mixing in the soft masses, the five
tadpole equations can be conveniently written in matrix form as
\begin{equation}
\left[t_u^0,t_d^0,t_1^0,t_2^0,t_3^0 \right]^T=
{\bf{M}^2_{tad}}\left[v_u,v_d,v_1,v_2,v_3\right]^T
\label{tadpoleMat}
\end{equation}
where the matrix ${\bf{M}^2_{tad}}$ is given in \cite{hirsch:2000ef}
and depends on the vevs only through the $D$ term defined above.

In the MSSM limit, where $\epsilon_i=v_i=0$, the angles $\theta_i$ are
equal to $\pi/2$.  In addition to the above MSSM parameters, our model
contains nine new parameters, $\epsilon_i$, $v_i$ and $B_i$. Considering 
we have three tadpole equations one can take either the 3 $B_i$ as 
input and derive the 3 sneutrino vevs or vice versa, such that we 
have in total just six new parameters (compared to the MSSM).

In order to have approximate solutions for the tree level vevs,
consider the following rotation among the $H_d$ and lepton
superfields:
\begin{equation}
{\bf{M'}^2_{tad}}={\bf R}{\bf{M}^2_{tad}}{\bf R^{-1}}
\label{rotMtad}
\end{equation}
where the rotation ${\bf R}$ can be split as
\begin{equation}
{\bf R}=
\left[
\begin{array}{ccccc}
c_3 & 0 &   0  &   0  & -s_3\\
 0  & 1 &   0  &   0  &   0\\
 0  & 0 &   1  &   0  &   0\\
 0  & 0 &   0  &   1  &   0 \\
  s_3 & 0 &   0  &   0  &  c_3\\
\end{array}
\right]
  \times
\left[
\begin{array}{ccccc}
c_2 & 0 &   0  & -s_2 &   0  \\
0  & 1 &   0  &   0  &   0  \\
0  & 0 &   1  &   0  &   0  \\
s_2 & 0 &   0  &  c_2 &   0  \\
0  & 0 &   0  &   0  &   1  \\
\end{array}
\right]
  \times
\left[
\begin{array}{ccccc}
c_1 & 0 & -s_1 &   0  &   0  \\
0  & 1 &   0  &   0  &   0  \\
s_1 & 0 &  c_1 &   0  &   0  \\
0  & 0 &   0  &   1  &   0  \\
0  & 0 &   0  &   0  &   1  \\
\end{array}
\right]\,.
\label{rotationR}
\end{equation}
where the three angles are defined as
\begin{eqnarray}
&&c_1={{\mu}\over{\mu'}}\,,\qquad\,\, s_1={{\epsilon_1}\over{\mu'}}\,,
\qquad\,\,\mu'=\sqrt{\mu^2+\epsilon_1^2}\,,\nonumber\\
&&c_2={{\mu'}\over{\mu''}}\,,\qquad\, s_2={{\epsilon_2}\over{\mu''}}\,,
\qquad\,\mu''=\sqrt{\mu'^2+\epsilon_2^2}\,,\label{angles}\\
&&c_3={{\mu''}\over{\mu'''}}\,,\qquad s_3={{\epsilon_3}\over{\mu'''}}\,,
\qquad\mu'''=\sqrt{\mu''^2+\epsilon_3^2}\,.\nonumber
\end{eqnarray}
It is clear that this rotation ${\bf R}$ leaves the $D$ term
invariant.  The rotated vevs are given by
\begin{equation}
\left[v'_u,v'_d,v'_1,v'_2,v'_3\right]^T=
{\bf{R}}\left[v_u,v_d,v_1,v_2,v_3\right]^T\,,
\label{rotatedvevs}
\end{equation}
and under the assumption that $v'_1,v'_2,v'_3\ll v$, these three small
vevs have the approximate solution
\begin{eqnarray}
&&v'_1\approx-{{\mu\epsilon_1}\over{M'^2_{L_1}+D}}\left[
{{m_{H_d}^2-M_{L_1}^2}\over{\mu'\mu'''}}v'_d+{{B_1-B}\over{\mu'}}v'_u
\right]\,,\nonumber\\\nonumber\\
&&v'_2\approx-{{\mu'\epsilon_2}\over{M'^2_{L_2}+D}}\left[
{{m'^2_{H_d}-M_{L_2}^2}\over{\mu''\mu'''}}v'_d+{{B_2-B'}\over{\mu''}}v'_u
\right]\,,\label{vevsPapprox}\\\nonumber\\
&&v'_3\approx-{{\mu''\epsilon_3}\over{M'^2_{L_3}+D}}\left[
{{m''^2_{H_d}-M_{L_3}^2}\over{\mu'''^2}}v'_d+{{B_3-B''}\over{\mu'''}}v'_u
\right]\,,\nonumber
\end{eqnarray}
where we have defined the following rotated soft terms:
\begin{eqnarray}
&&\!\!\!\!\!\!\!\!\!\!\!\!\!\!\!\!\!m'^2_{H_d}=
{{m_{H_d}^2\mu^2+M_{L_1}^2\epsilon_1^2}\over{\mu'^2}}\,,\quad
m''^2_{H_d}={{m'^2_{H_d}\mu'^2+M_{L_2}^2\epsilon_2^2}\over{\mu''^2}}
\,,\quad
m'''^2_{H_d}={{m''^2_{H_d}\mu''^2+M_{L_3}^2\epsilon_3^2}\over{\mu'''^2}}
\,,\nonumber\\
&&\!\!\!\!\!\!\!\!\!\!\!\!\!\!\!\!\!B'=
{{B\mu^2+B_1\epsilon_1^2}\over{\mu'^2}}\,,\qquad\quad\,\,\,
B''={{B'\mu'^2+B_2\epsilon_2^2}\over{\mu''^2}}\,,\qquad\quad
B'''={{B''\mu''^2+B_3\epsilon_3^2}\over{\mu'''^2}}\,,
\label{newsoft}\\
&&\!\!\!\!\!\!\!\!\!\!\!\!\!\!\!\!\!M'^2_{L_1}=
{{m_{H_d}^2\epsilon_1^2+M_{L_1}^2\mu^2}\over{\mu'^2}}\,,\quad
M'^2_{L_2}={{m'^2_{H_d}\epsilon_2^2+M_{L_2}^2\mu'^2}\over{\mu''^2}}\,,
\quad
M'^2_{L_3}={{m''^2_{H_d}\epsilon_3^2+M_{L_3}^2\mu''^2}\over{\mu'''^2}}\,.
\nonumber
\end{eqnarray}
As can be seen from eq. (17) the approximation $v'_1,v'_2,v'_3\ll v$
is justified if either a) $\epsilon_i \ll \mu$ and/or b)
$(m_{H_d}^2-M_{L_i}^2)/\mu^2\ll 1$ and $(B_i-B)/\mu \ll 1$.  The
latter holds automatically (to some extent) in many models of
supersymmetry breaking, as for example in minimal supergravity
Ref.~\cite{diaz:1998xc}.

As in the MSSM, the electroweak symmetry is broken because the large
value of the top quark mass drives the Higgs mass parameter
$m_{H_U}^2$ to negative values at the weak scale via its
RGE~\cite{Ibanez:1982fr}. In the rotated basis, the parameter
$\mu'''^2$ is determined at one loop by
\begin{equation}
\mu'''^2=-{1\over 2}\left[m_Z^2-\widetilde A_{ZZ}(m_Z^2)\right]+
{{\left(m'''^2_{H_d}+\widetilde T^{\overline{DR}}_{v'_d}\,\right)-
\left(m_{H_u}^2+\widetilde T^{\overline{DR}}_{v'_u}\,\right)t'^2_{\beta}
}\over{t'^2_{\beta}-1}}
\label{muppp2}
\end{equation}
where $t'_{\beta}=v'_u/v'_d$ is defined in the rotated basis and is
analogous to $\tan\beta$ in eq.~(\ref{eq:vevs}) defined in the
original basis. The finite $\overline{DR}$ Z-boson self energy is
$\widetilde A_{ZZ}(m_Z^2)$, and the one--loop tadpoles
$T^{\overline{DR}}_{v'_d}$ and $T^{\overline{DR}}_{v'_u}$ are obtained
by applying to the original tadpoles in eq.~(\ref{tadpoles}) the
rotation $\bf R$ defined in eq.~(\ref{rotationR}).  The radiative
breaking of the electroweak symmetry is valid in the BRpV model in the
usual way: the large value of the top quark Yukawa coupling drives the
parameter $m_{H_U}^2$ to negative values, breaking the symmetry of the
scalar potential.

\subsection{Neutral fermion mass matrix}
\label{sec:neutral-fermion-mass}

Here we consider the tree level structure of the fermion mass matrices
in this model. For a more complete discussion of different mass matrices 
in BRpV see the Appendix of Ref.~\cite{hirsch:2000ef}.
In the basis $\psi^{0T}=
(-i\lambda',-i\lambda^3,\widetilde{H}_d^1,\widetilde{H}_u^2, \nu_{e},
\nu_{\mu}, \nu_{\tau} )$ the neutral fermion mass matrix ${\bold M}_N$
is given by
\begin{equation}
{\bold M}_N=\left[  
\begin{array}{cc}  
{\cal M}_{\chi^0}& m^T \cr
\vb{20}
m & 0 \cr
\end{array}
\right]
\end{equation}
where
\begin{equation}
{\cal M}_{\chi^0}\hskip -2pt=\hskip -4pt \left[ \hskip -7pt 
\begin{array}{cccc}  
M_1 & 0 & -\frac 12g^{\prime }v_d & \frac 12g^{\prime }v_u \cr
\vb{12}   
0 & M_2 & \frac 12gv_d & -\frac 12gv_u \cr
\vb{12}   
-\frac 12g^{\prime }v_d & \frac 12gv_d & 0 & -\mu  \cr
\vb{12}
\frac 12g^{\prime }v_u & -\frac 12gv_u & -\mu & 0  \cr
\end{array}  
\hskip -6pt
\right] 
\end{equation}
is the standard MSSM neutralino mass matrix ($M_2$ and $M_1$ are the
SU(2) and U(1) gaugino soft masses) and
\begin{equation}
m=\left[  
\begin{array}{cccc}  
-\frac 12g^{\prime }v_1 & \frac 12gv_1 & 0 & \epsilon _1 \cr
\vb{18}
-\frac 12g^{\prime }v_2 & \frac 12gv_2 & 0 & \epsilon _2  \cr
\vb{18}
-\frac 12g^{\prime }v_3 & \frac 12gv_3 & 0 & \epsilon _3  \cr  
\end{array}  
\right] 
\end{equation}
characterizes the breaking of R-parity.  The full $7\times 7$
neutrino/neutralino mass matrix ${\bold M}_N$ is diagonalized as
\begin{equation}
{\cal  N}^*{\bold M}_N{\cal N}^{-1}={\rm diag}(m_{\chi^0_i},m_{\nu_j})
\label{chi0massmat}
\end{equation}
where $(i=1,\cdots,4)$ for the neutralinos, and $(j=1,\cdots,3)$ for
the neutrinos.


%
\begin{equation}
{\cal N}^* {\bf M}_{F^0} {\cal N}^{-1}={\bf M}_{F^0}^{\mathrm diag}
\end{equation}
and the eigenvectors are given by
\begin{equation}
F^0_i={\cal N}_{ij}\psi_j
\end{equation}
using the basis $\psi=(-i\lambda',-i\lambda^3,\widetilde{H}_d^1,
\widetilde{H}_u^2, \nu_{e},\nu_{\mu}, \nu_{\tau} )$.  As discussed in
more detail below, to a very good approximation, the rotation matrix
can be written as
\begin{equation}
{\cal N}^*\approx\left(
  \begin{array}{cc}
N^* & N^*\xi^{\dagger} \\
-V_{\nu}^T\xi & V_{\nu}^T 
  \end{array}
\right)
\end{equation}
Here, $N$ is the rotation matrix that diagonalizes the $4\times4$ MSSM
neutralino mass matrix, $V_{\nu}$ is the rotation matrix that
diagonalizes the tree level neutrino $3\times3$ mass matrix, and
$\xi_{ij}\ll 1$ are the relevant small expansion parameters which
characterize the violation of R--parity and whose form will be given
in Sec.~\ref{sec:tree-level-mass}.

\subsection{Charged fermion mass matrix}
\label{sec:charged-fermion-mass}

The chargino/lepton mass matrix is given by 

\begin{equation} 
{\bold M_C}=\left[
\begin{array}{ccccc} 
M_2 & {\textstyle{1\over{\sqrt{2}}}}gv_u & 0 &0& 0\\ 
\vb{18}
{\textstyle{1\over{\sqrt{2}}}}gv_d & \mu &  
-{\textstyle{1\over{\sqrt{2}}}}\left(h_E\right)_{11}v_1& 
-{\textstyle{1\over{\sqrt{2}}}}\left(h_E\right)_{22}v_2& 
-{\textstyle{1\over{\sqrt{2}}}}\left(h_E\right)_{33}v_3 \\
\vb{18} 
{\textstyle{1\over{\sqrt{2}}}}gv_1 & -\epsilon_1 & 
{\textstyle{1\over{\sqrt{2}}}}\left(h_E\right)_{11}v_d&0&0\\
{\textstyle{1\over{\sqrt{2}}}}gv_2 & -\epsilon_2 & 0&
{\textstyle{1\over{\sqrt{2}}}}\left(h_E\right)_{22}v_d&0\\
{\textstyle{1\over{\sqrt{2}}}}gv_3 & -\epsilon_3 & 0&0&
{\textstyle{1\over{\sqrt{2}}}}\left(h_E\right)_{33}v_d
\end{array}
\right] 
\label{eq:ChaM6x6} 
\end{equation} 
We note that the chargino sector decouples from the lepton sector in
the limit $\epsilon_i=v_i=0$. As in the MSSM, the chargino mass matrix
is diagonalized by two rotation matrices $\bold U$ and $\bold V$
defined by

\begin{equation}
{\cal U}^*\, {\bf M}_{F^+} {\cal V}^{-1}={\bf M}_{F^+}^{\mathrm diag}
\end{equation}
with the eigenvectors satisfying
\begin{equation}
F^+_{Ri}={\cal V}_{ij}\psi^+_j\,,\qquad 
F^-_{Li}={\cal U}_{ij}\psi^-_j
\end{equation}
in the basis 
$\psi^+=(-i\lambda^+,\widetilde H_2^1,e_R^+,\mu_R^+,\tau_R^+)$ and
$\psi^-=(-i\lambda^-,\widetilde H_1^2,e_L^-,\mu_L^-,\tau_L^-)$, and with
the Dirac fermions being
\begin{equation}
F_i^+=\left(\begin{array}{c}F^+_{Ri} \\ \\ \overline{F^-_{Li}}\end{array}\right)
\end{equation}
To first order in the R-Parity violating parameters we have
\begin{equation}
{\cal V}\approx\left(\begin{array}{cc}
V & V\xi^{T}_R \\
-V^{\ell}_R\xi^{*}_R & V^{\ell}_R
\end{array}\right)\,,\qquad
{\cal U}\approx\left(\begin{array}{cc}
U & U\xi^{\dagger}_L \\
-V^{{\ell}*}_L\xi_L & V^{{\ell}*}_L
\end{array}\right)
\end{equation}
where $V^{{\ell}*}_L$ and $V^{\ell}_R$ diagonalize the charged lepton
mass matrix according to $V^{{\ell}*}_L{\bf
  M}^{\ell}V^{\ell\dagger}_R={\bf M}^{\ell}_{\mathrm diag}$. For most 
purposes it is sufficient to take $\xi_R={\bf 0}_{2\times3}$, since 
it is smaller than $\xi_L$ typically by a factor of $m_l/m_{SUSY}$. 
Note, that we can choose $V^{{\ell}*}_L = 
V^{\ell\dagger}_R = {\bf 1}_{3\times 3}$. We then have
\begin{equation}
\xi_L^{i1}=a^L_1\Lambda_i\,,\qquad \xi_L^{i2}=a^L_2\Lambda_i+b\epsilon_i
\end{equation}
and
\begin{equation}
a_1^L={g\over{\sqrt{2}\Delta_+}}\,,\qquad 
a_2^L=-{{g^2v_u}\over{2\mu\Delta_+}}
\end{equation}
where $\Delta_+$ is the determinant of the $2\times2$ chargino mass matrix 
and 
\begin{equation}
\Lambda_i = \mu v_i + v_d \epsilon_i \propto  v'_i
\label{lambdai}
\end{equation}
are the alignment parameters. 

\section{Tree--level neutrino mass: the atmospheric scale}
\label{sec:tree-level-mass}

The tree-level contribution to neutrino masses from broken R parity
supersymmetry has a long history~\cite{santamaria:1987uq}. Thanks to
the Super-K findings~\cite{fukuda:1998mi} we will be interested only
in the case where the neutrino mass which is determined at the tree
level is small, in order to account for the atmospheric neutrino data.
The above form for ${\bold M}_N$ is especially convenient in this case
in order to provide an approximate analytical discussion valid in the
limit of small \rp violation parameters. Indeed in this case we
perform a perturbative diagonalization of the neutral mass matrix,
defining
\begin{equation}
\xi = m \cdot {\cal M}_{\chi^0}^{-1}
\label{defxi}
\end{equation}
Since the effective RpV parameters are smaller than the weak scale, we
can work in a perturbative expansion defined by $\xi \ll 1$, where
$\xi$ denotes a $3\times 4$ matrix given as~\cite{Hirsch:1998kc}
\begin{eqnarray}
\xi_{i1} &=& \frac{g' M_2 \mu}{2 det({\cal M}_{\chi^0})}\Lambda_i \cr
\vb{20}
\xi_{i2} &=& -\frac{g M_1 \mu}{2 det({\cal M}_{\chi^0})}\Lambda_i \cr
\vb{20}
\xi_{i3} &=& - \frac{\epsilon_i}{\mu} + 
          \frac{(g^2 M_1 + {g'}^2 M_2) v_u}
               {4 det({\cal M}_{\chi^0})}\Lambda_i \cr
\vb{20}
\xi_{i4} &=& - \frac{(g^2 M_1 + {g'}^2 M_2) v_d}
               {4 det({\cal M}_{\chi^0})}\Lambda_i
\label{xielementos}
\end{eqnarray}
From Eq.~(\ref{xielementos}) and Eq.~(\ref{lambdai}) one can see that
$\xi=0$ in the MSSM limit where $\epsilon_i=0$, $v_i=0$.

If the elements of this matrix satisfy
\begin{equation}
\forall \xi_{ij} \ll 1
\end{equation}
then one can use it as expansion parameter in order to find an
approximate solution for the mixing matrix ${\cal N}$.  

In leading order in $\xi$ the mixing matrix ${\cal N}$ is
given by,
\begin{equation}
{\cal N}^* \hskip -2pt =\hskip -2pt  \left(\hskip -1pt
\begin{array}{cc}
N^* & 0\\
0& V_\nu^T \end{array}
\hskip -1pt
\right)
\left(
\hskip -1pt
\begin{array}{cc}
1 -{1 \over 2} \xi^{\dagger} \xi& \xi^{\dagger} \\
-\xi &  1 -{1 \over 2} \xi \xi^\dagger
\end{array}
\hskip -1pt
\right) 
\end{equation}
The second matrix above block-diagonalizes the mass matrix ${\bold
  M}_N$ approximately to the form diag(${\cal M}_{\chi^0},m_{eff}$),
where
\begin{eqnarray}
m_{eff} &=& 
- m \cdot {\cal M}_{\chi^0}^{-1} m^T \cr
\vb{18}
&\hskip -3mm=\hskip -3mm& 
\frac{M_1 g^2 \!+\! M_2 {g'}^2}{4\, det({\cal M}_{\chi^0})} 
\left(\hskip -2mm \begin{array}{ccc}
\Lambda_e^2 
\hskip -1pt&\hskip -1pt
\Lambda_e \Lambda_\mu
\hskip -1pt&\hskip -1pt
\Lambda_e \Lambda_\tau \\
\Lambda_e \Lambda_\mu 
\hskip -1pt&\hskip -1pt
\Lambda_\mu^2
\hskip -1pt&\hskip -1pt
\Lambda_\mu \Lambda_\tau \\
\Lambda_e \Lambda_\tau 
\hskip -1pt&\hskip -1pt 
\Lambda_\mu \Lambda_\tau 
\hskip -1pt&\hskip -1pt
\Lambda_\tau^2
\end{array}\hskip -3mm \right)
\end{eqnarray}

\noindent
The sub-matrices $N$ and $V_{\nu}$ diagonalize ${\cal M}_{\chi^0}$ and
$m_{eff}$
\begin{equation}
N^{*}{\cal M}_{\chi^0} N^{\dagger} = {\rm diag}(m_{\chi^0_i}),
\end{equation}
\begin{equation}
V_{\nu}^T m_{eff} V_{\nu} = {\rm diag}(0,0,m_{\nu}),
\end{equation}
where 
\begin{equation}
\label{mnutree}
m_{\nu} = Tr(m_{eff}) = 
\frac{M_1 g^2 + M_2 {g'}^2}{4\, det({\cal M}_{\chi^0})} 
|{\vec \Lambda}|^2.
\end{equation}
The special form of the neutralino/neutrino mass matrix implies that
the effective neutrino mass matrix $ m_{eff}$ generated after
diagonalizing out the heavy neutralinos has a projective form, a
feature common to many spontaneous R--parity violating models
\cite{santamaria:1987uq}.  This implies that only one neutrino
acquires a tree level mass, the other two remaining
massless~\cite{santamaria:1987uq}.  As a result at the tree
approximation one can rotate away one of the three angles
in the matrix $V_{\nu}$, leading to
\begin{equation}
V_{\nu}= 
\left(\begin{array}{ccc}
  1 &                0 &               0 \\
  0 &  \cos\theta_{23} & -\sin\theta_{23} \\
  0 &  \sin\theta_{23} & \cos\theta_{23} 
\end{array}\right) \times 
\left(\begin{array}{ccc}
  \cos\theta_{13} & 0 & -\sin\theta_{13} \\
                0 & 1 &               0 \\
  \sin\theta_{13} & 0 & \cos\theta_{13} 
\end{array}\right) ,
\end{equation}
where the mixing angles can be expressed in terms of the {\it
  alignment vector} ${\vec \Lambda}$ as follows:
\begin{equation}
\label{tetachooz}
\tan\theta_{13} = - \frac{\Lambda_e}
                   {(\Lambda_{\mu}^2+\Lambda_{\tau}^2)^{\frac{1}{2}}},
\end{equation}
\begin{equation}
\label{tetatm}
\tan\theta_{23} = - \frac{\Lambda_{\mu}}{\Lambda_{\tau}}.
\end{equation}
The non-zero tree--level eigenvalue of the neutrino mass matrix is
identified with the atmospheric mass scale.  
\begin{figure}[h]
  \centering
      \includegraphics[height=5cm,width=0.8\linewidth]{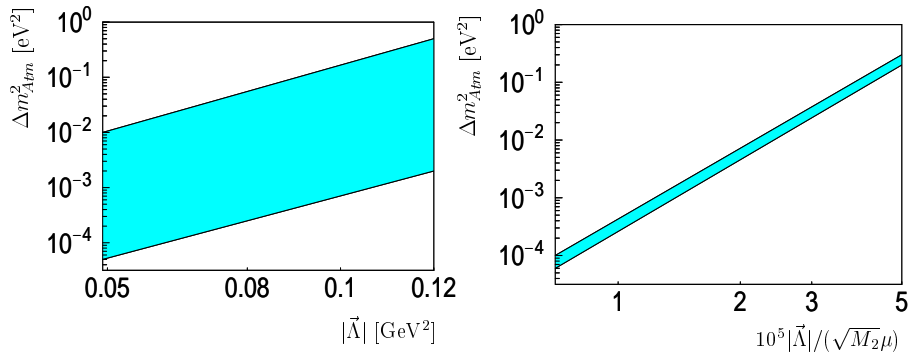}
  \caption{$\Delta m^2_{atm}$  versus the  BRpV alignment 
    parameters }
  \label{fig:atmmass}
\end{figure}
The calculated $\Delta m^2_{atm}$ can be expressed as a function of
the alignment parameter ${\vec \Lambda}$ (left in
Fig.~\ref{fig:atmmass}), or as function of $|{\vec
  \Lambda}|/(\sqrt{M_2} \mu)$ (right in Fig.~\ref{fig:atmmass}), all
of these expressed in GeV.  The figure shows that Eq.~(\ref{mnutree})
can be used to fix the relative size of R-parity breaking parameters
to obtain the correct $\Delta m^2_{atm}$.
On the other hand, as shown in Fig~~\ref{fig:atmangle} the atmospheric
angle can be expressed in terms of
$\Lambda_{\mu}/\Lambda_{\tau}$. Its maximality
is obtained for $\Lambda_{\mu} \simeq \Lambda_{\tau}$ if $\Lambda_{e}$
is smaller than the other two.
\begin{figure}[htbp]
  \centering
      \includegraphics[height=6cm,width=0.7\linewidth]{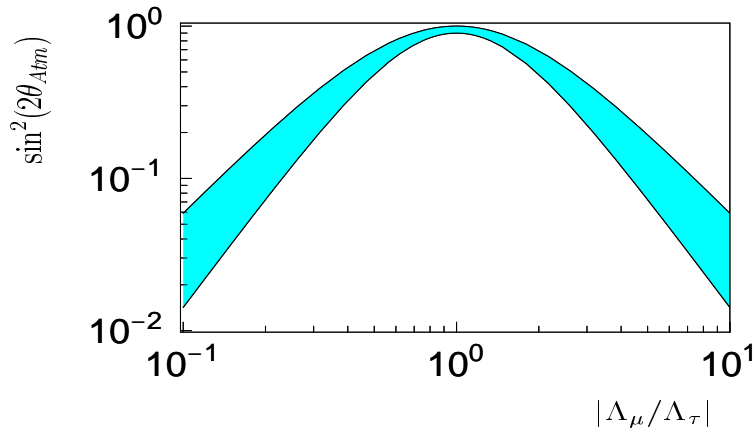}
  \caption{The atmospheric angle versus the ratio of BRpV parameters 
$|\Lambda_{\mu}/\Lambda_{\tau}|$}
  \label{fig:atmangle}
\end{figure}
Let us stress once again that there is no solar mass splitting in the
tree approximation so that, as a result, the ``solar angle'' is not
defined, as it can be rotated away by redefining the two degenerate
neutrinos~\cite{schechter:1980gr}.

\section{One--loop--induced neutrino mass: the solar scale }
\label{sec:one-loop-induced}

As we just saw in the BRpV model the atmospheric mass scale and mixing
arises at the tree-level.  We now discuss the determination of solar 
neutrino masses and mixings, which are both generated radiatively. 
One-loop corrections to the neutrino masses can be calculated 
numerically \cite{hirsch:2000ef} or analytically \cite{Diaz:2003as}. 
While the numerical approach can give ``exact'' results (exact in 
the sense of being correct up to higher order effects), the analytic 
approach, while being less accurate, gives a better understanding 
about which parameters control the loops and thus the solar neutrino 
mass and angles in our model. The discussion will therefore mainly 
concentrate on the analytical calculations.

In principle, in order to find the correct neutrino mixing angles one 
has to diagonalize the one--loop corrected neutralino/neutrino mass 
matrix. We define
\begin{equation}
M^{\rm pole}_{ij}= M^{\rm \overline{DR}}_{ij}(Q) + \Delta M_{ij}
\end{equation}
where $M^{\rm \overline{DR}}_{ij}(Q)$ is the tree-level pole mass 
and ${\rm \overline{DR}}$ indicates the dimensional reduction scheme 
we used in the numerical calculation. One-loop corrections are 
\begin{equation}
\Delta M_{ij}= \half 
\left[\widetilde\Pi^V_{ij}(m_i^2) + \widetilde\Pi^V_{ij}(m_j^2)\right] 
- \half 
\left[ m_{\chi^0_i} \widetilde\Sigma^V_{ij}(m_i^2)  +
m_{\chi^0_j} \widetilde\Sigma^V_{ij}(m_j^2) \right]\,,
\label{DeltaM}
\end{equation}
where the symmetrization is necessary to achieve gauge invariance and
consistency with the Pauli principle.  Here $\widetilde\Pi^V$ and
$\widetilde\Sigma^V$ are the renormalized self-energies. They contain
products of couplings and the usual Passarino-Veltman functions
\cite{passarino:1979jh}.

Diagonalizing the tree-level neutrino mass matrix first and adding
then the one-loop corrections before re-diagonalization one finds that
the resulting neutrino/neutralino mass matrix has non-zero entries in
the neutrino/neutrino, the neutrino/neutralino and in the
neutralino/neutralino sectors. We have found \cite{Diaz:2003as} that 
the most important part of the one-loop neutrino masses derives from the
neutrino/neutrino sector and that the one-loop induced
neutrino/neutralino mixing is usually negligible.

\begin{figure}[ht]
  \begin{center}
  \begin{tabular}{ccc}
    \includegraphics[width=0.28\linewidth]{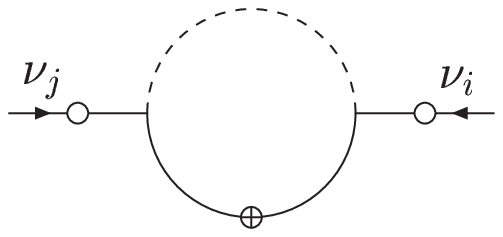}&
    \includegraphics[width=0.28\linewidth]{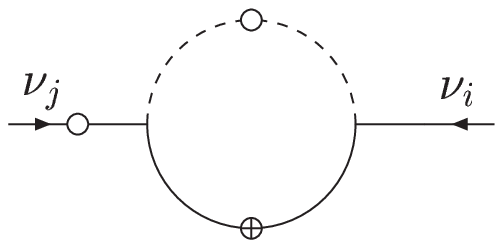}&
    \includegraphics[width=0.28\linewidth]{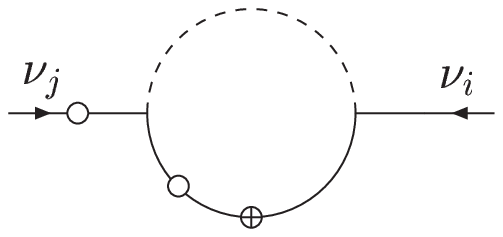}
  \end{tabular}
  \begin{tabular}{ccc}
    \includegraphics[width=0.28\linewidth]{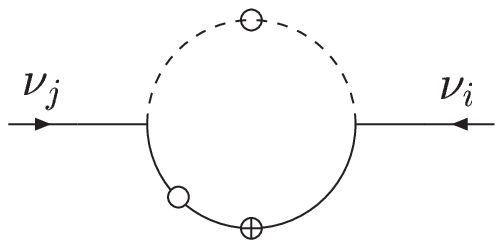}&
     & \includegraphics[bb= 115 628 260 678,height=15mm]{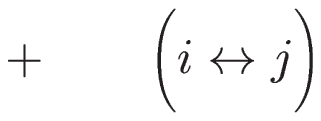}
  \end{tabular}
\caption{Topologies for neutrino self-energies in the BRpV 
supersymmetric model}
\label{fig:topologies}
\end{center}
\end{figure}

The relevant topologies for the one loop calculation of neutrino
masses are then illustrated in Fig.\ref{fig:topologies}.
Here our conventions are as follows: open circles with a cross inside
indicate genuine mass insertions which flip chirality.  On the other
hand open circles without a cross correspond to small R-Parity
violating projections, indicating how much of an Rp-even/odd mass
eigenstate is present in a given Rp-odd/even weak eigenstate.  In the
actual numerical calculation these projections really belong to the
coupling matrices attached to the vertices.  However, given the
smallness of Rp-violating effects, the pre-diagonalization
``insertion-method'' proves to be a rather useful tool to develop an
analytical perturbative expansion and to acquire a simple
understanding of the results.

These topologies have then to be ``filled'' with all relevant 
combinations of particles/sparticles. Here we will concentrate on 
discussing the loop involving bottom/sbottom quarks, since a) it is 
in large  parts of parameter space the numerically most important one. 
And b) other loops can be calculated in a very similar way \cite{Diaz:2003as}, 
although they are more complicated.

The relevant Feynman rules for the bottom-sbottom loops are, in the
case of left sbottoms:
\begin{center}
\includegraphics{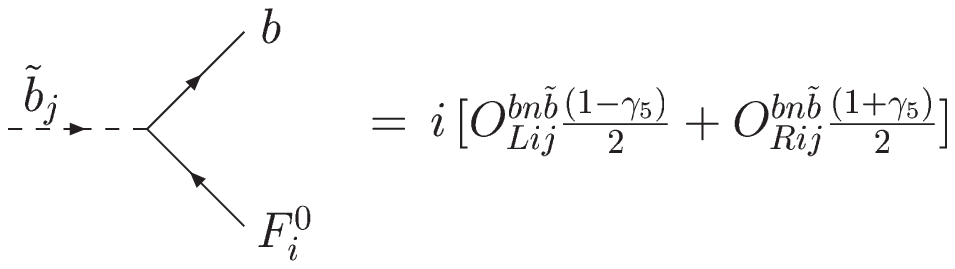}
\end{center}
with
\begin{eqnarray}
O^{bn\tilde b}_{Lij}&=&-R^{\tilde b}_{j1}h_b{\cal N}^*_{i3}
-R^{\tilde b}_{j2}{{2g}\over{3\sqrt{2}}}\tan\theta_W{\cal N}^*_{i1}
\nonumber\\
O^{bn\tilde b}_{Rij}&=&R^{\tilde b}_{j1}{g\over{\sqrt{2}}}
\left({\cal N}_{i2}-{\textstyle{1\over3}}\tan\theta_W{\cal N}_{i1}\right)
-R^{\tilde b}_{j2}h_b{\cal N}^*_{i3}
\label{Obnsb}
\end{eqnarray}
After approximating the rotation matrix ${\cal N}$ we find that
expressions with the replacement ${\cal
  N}\to {\mathrm N}$ are valid when the neutral fermion is a
neutralino. When the neutral fermion $F^0$ is a neutrino, the
following expressions hold
\begin{eqnarray}
O^{bn\tilde b}_{Lij} &\approx& R^{\tilde b}_{j1}h_b\left(
a_3|\vec\Lambda|\delta_{i'3}+b\tilde\epsilon_{i'}\right)
+R^{\tilde b}_{j2}{{2g}\over{3\sqrt{2}}}\tan\theta_W a_1|\vec\Lambda|\delta_{i'3}
\nonumber\\
O^{bn\tilde b}_{Rij} &\approx& R^{\tilde b}_{j1}{g\over{\sqrt{2}}}
\left({\textstyle{1\over3}}\tan\theta_W a_1-a_2\right)|\vec\Lambda|\delta_{i'3}
+R^{\tilde b}_{j2}h_b\left(a_3|\vec\Lambda|\delta_{i'3}+
b\tilde\epsilon_{i'}\right)
\end{eqnarray}
where $i'=i-4$ label one of the neutrinos. $R^{\tilde b}_{jk}$ are the 
rotation matrices connecting weak and mass eigenstate basis for the 
scalar bottom quarks. In case of no intergenerational mixing in the 
squark sector $R^{\tilde b}_{jk}$ can be parameterized by just one 
diagonalizing angle $\theta_{\tilde b}$.

\begin{figure}[ht]
\begin{center}
\begin{tabular}{cc}
\includegraphics[width=0.45\linewidth]{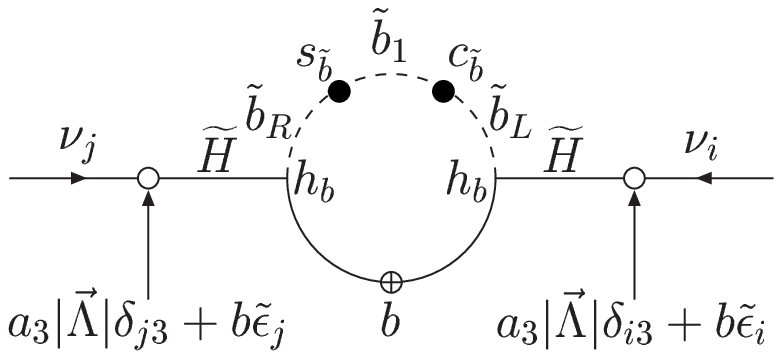}
&\includegraphics[width=0.45\linewidth]{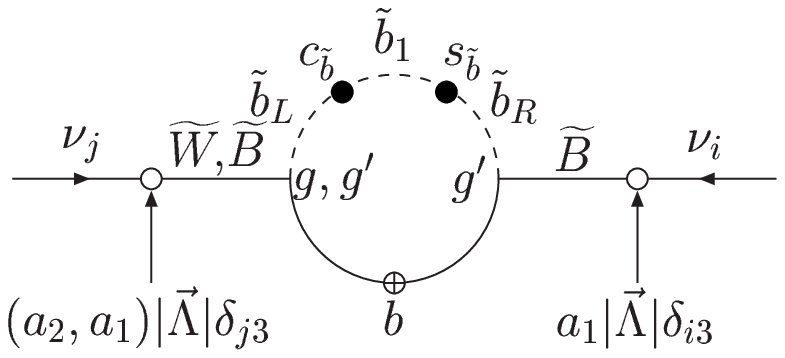}
\end{tabular}
\end{center}
\vspace{-3mm}
\caption{Bottom--Sbottom diagrams for solar neutrino mass in the BRpV model}
\label{fig:BottomSbottomLoop}
\end{figure}

Putting these couplings together one finds 
the simplest contribution to the radiatively induced neutrino mass
from loops involving bottom quarks and squarks 
\cite{hirsch:2000ef}
\begin{equation}
\widetilde\Pi_{ij}(0)=-{{N_c}\over{16\pi^2}}\sum_r\left(
O^{bn\tilde b}_{Rjr}O^{bn\tilde b}_{Lir}+
O^{bn\tilde b}_{Ljr}O^{bn\tilde b}_{Rir}\right)
m_bB_0(0,m_b^2,m_r^2)
\end{equation}
where $B_0(0,m_b^2,m_r^2)$ is a Passarino-Veltman function
\cite{passarino:1979jh} can be written as follows
\begin{eqnarray}
\label{eq:bsb}
\widetilde\Pi_{ij}&=&-{{N_c m_b}\over{16\pi^2}}
2s_{\tilde b}c_{\tilde b}h_b^2
\Delta B_0^{\tilde b_1\tilde b_2} \times \\ \nonumber
& & \Big(
 {{\tilde\epsilon_i\tilde\epsilon_j}\over{\mu^2}}
  + a_3 b \left(\tilde\epsilon_i\delta_{j3}+
 \tilde\epsilon_j\delta_{i3}\right)|\vec\Lambda|
 + \left( a_3^2 + \frac{ a_L a_R}{h^2_b}\right)
 \delta_{i3}\delta_{j3} 
 |\vec\Lambda|^2 \Big) 
 \end{eqnarray}
This expression is proportional to the
difference of two $B_0$ functions,
\begin{equation}
  \label{eq:PV}
  \Delta B_0^{\tilde b_1\tilde b_2}=
  B_0(0,m_b^2,m_{\tilde b_1}^2)-B_0(0,m_b^2,m_{\tilde b_2}^2)
\end{equation}
Parameters $\Lambda_i$ have been defined above. 
The $\tilde\epsilon$ parameters are defined as
 $\tilde\epsilon_i=\left(V_{\nu}^T\right)^{ij}\epsilon_j$, and are
 given by
\begin{eqnarray}
\label{eq:TildeEpsilon}
\tilde\epsilon_1&=&{{
\epsilon_e(\Lambda_{\mu}^2+\Lambda_{\tau}^2)-\Lambda_e
(\Lambda_{\mu}\epsilon_{\mu}+\Lambda_{\tau}\epsilon_{\tau})
}\over{
\sqrt{\Lambda_{\mu}^2+\Lambda_{\tau}^2}
\sqrt{\Lambda_e^2+\Lambda_{\mu}^2+\Lambda_{\tau}^2}
}}
\nonumber\\
\tilde\epsilon_2&=&{{
\Lambda_{\tau}\epsilon_{\mu}-\Lambda_{\mu}\epsilon_{\tau}
}\over{
\sqrt{\Lambda_{\mu}^2+\Lambda_{\tau}^2}
}}
\\
\tilde\epsilon_3&=&{{
\vec\Lambda\cdot\vec\epsilon
}\over{
\sqrt{\Lambda_e^2+\Lambda_{\mu}^2+\Lambda_{\tau}^2}
}}\nonumber
\end{eqnarray}
On the other hand $a_{L,R}$ are defined as
\begin{equation}
a_R={g\over{\sqrt{2}}}\left({1\over3}t_Wa_1-a_2\right)
\,,\qquad
a_L={g\over{\sqrt{2}}}\,{2\over3}t_Wa_1
\end{equation}
The different terms in eq. (\ref{eq:bsb}) can be understood as coming from the
graphs corresponding to the first topology of
Fig.~\ref{fig:topologies}. They have been depicted in more detail in
Fig.~\ref{fig:BottomSbottomLoop},
where we have adopted the following conventions: a) as before, open
circles correspond to small R-parity violating projections, indicating
how much of a weak eigenstate is present in a given mass eigenstate,
(b) full circles correspond to R-parity conserving projections and (c)
open circles with a cross inside indicate genuine mass insertions
which flip chirality.

The open and full circles should really appear at the vertices since
the particles propagating in the loop are the mass eigenstates. We
have however separated them to better identify the origin of the
various terms.
There is another set of graphs analogous to the previous ones which
corresponds to the heavy sbottom. They are obtained from the previous
graphs making the replacement $\tilde b_1\to \tilde b_2$,
$s_{\tilde b}\to  c_{\tilde b}$ and $c_{\tilde b}\to 
-s_{\tilde b}$.  Note that for all contributions to the $2\times2$
sub-matrix corresponding to the light neutrinos the divergence from
$B_0(0,m_b^2,m_{\tilde b_1}^2)$ is canceled by the divergence from
$B_0(0,m_b^2,m_{\tilde b_2}^2)$, making finite the contribution from
bottom-sbottom loops to this sub-matrix, as it should be, since the
mass is fully ``calculable''.

The second most important contribution to the radiatively induced neutrino 
mass usually comes from charged-scalar/charged-fermion 
loops \cite{hirsch:2000ef}. Since all possible topologies of Fig. (3) 
contribute to this loop the structure of the contribution from charged 
Higgs/slepton loops is more complex than that of the bottom-sbottom loop
considered above. However, the same topology as for the sbottom/bottom 
loop also contributes to the charged scalar loop. It leads to a final 
expression similar to eq. (\ref{eq:bsb}), with appropriate replacements, 
which is good enough for an order-of-magnitude estimate of the 
charged scalar loop.

\subsection{Results for the solar mass scale}
\label{sec:analyt-appr-solar-1}

\begin{figure}
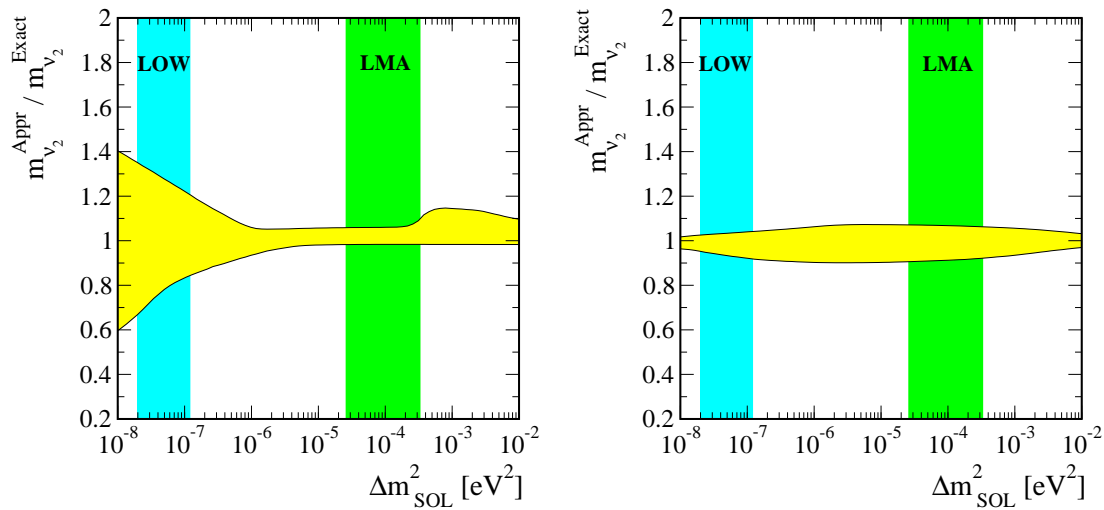
\centering
  \begin{tabular}{cc}
  \includegraphics[width=0.45\linewidth]{NfDelM-v4.eps}
  &\includegraphics[width=0.45\linewidth]{Stau4aDelM-v4.eps}
  \end{tabular}
\caption[] {($m_{\nu_2}^{Appr}/m_{\nu_2}^{exact}$) versus 
  $\Dms$ [$eV^2$] for the set Ntrl (left) and the set Stau (right). 
  $m_{\nu_2}^{Appr}$ is the sum of
  the bottom-sbottom and charged scalar loops, while
  $m_{\nu_2}^{exact}$ is the numerical result for all loops. In
  case of LMA the approximation works always better than 10 \%. }
\label{fig:Stau}
\end{figure}

We give a discussion of the analytical versus numerical results of 
the solar mass scale first. In Fig. (\ref{fig:Stau}) we show a 
comparison of approximate and exact calculation for two different 
numerical data sets. In both figures we show the ratio of the 
approximate-over-exact solar
neutrino mass parameter $m_{\nu_2}^{Appr}/m_{\nu_2}^{exact}$ versus
$\Dms$ in eV$^2$, where $m_{\nu_2}^{Appr}$ is the approximate loop
calculation involving the bottom-sbottom and the charged scalar loop,
while $m_{\nu_2}^{exact}$ is the exact numerical computation taking
into account all loops. The set to the left called ``Ntrl'' contains 
neutralinos being the LSP, while the set to the right (Stau) has the 
charged scalar tau as LSP.

We have found numerically that the terms proportional to ${\tilde
  \epsilon_i}\times {\tilde \epsilon_j}$ in the self energies in
Eq.~(\ref{eq:bsb}) give the most important contribution to $m_{\nu_2}$
in the bottom-sbottom loop calculation in most points of our sets. If
these terms are dominant one can find a very simple approximation for
the bottom-sbottom loop contribution to $m_{\nu_2}$.  It is given by
\begin{equation}
\label{Simplest}
m_{\nu_2} \simeq \frac{3}{16 \pi^2} \sin(2\theta_{\tilde b}) 
m_b \Delta B_0^{\tilde b_2\tilde b_1}\ 
\frac{({\tilde \epsilon}_1^2 + {\tilde \epsilon}_2^2)}{\mu^2}.
\end{equation}
Eq.~(\ref{Simplest}) works surprisingly well for almost all points 
in our data sets.

The more complicated structure of the charged scalar loop makes it
difficult to give a simple equation for $m_{\nu_2}$ similar to
Eq.~(\ref{Simplest}) for the bottom-sbottom loop.  However, we
note that Eq.  (\ref{Simplest}), with appropriate
replacements, allows us to estimate the typical contributions to the
charged scalar loop within a factor of $\sim 3$.  However, such an
estimate will be biased toward too small or too large $m_{\nu_2}$ 
depending mainly on which SUSY particle is the LSP \cite{Diaz:2003as}.

\subsection{Analytical approximation for the solar mixing angle}
\label{sec:analyt-appr-solar}

In the basis where the tree-level neutrino mass matrix is diagonal the
mass matrix at one--loop level can be written as
\begin{eqnarray}
  \label{eq:2}
  \widetilde m_\nu= V_\nu^{(0)T} m_\nu V_\nu^{(0)} 
      =\left( \begin{array}{ccc}
      c_1 \widetilde \epsilon_1 \widetilde \epsilon_1 & c_1 \widetilde
      \epsilon_1 \widetilde \epsilon_2  
      & c_1 \widetilde \epsilon_1 \widetilde \epsilon_3 \\
      c_1 \widetilde \epsilon_2 \widetilde \epsilon_1 & c_1 \widetilde
      \epsilon_2 \widetilde \epsilon_2  
      & c_1 \widetilde \epsilon_2 \widetilde \epsilon_3 \\
      c_1 \widetilde \epsilon_3 \widetilde \epsilon_1 & c_1 \widetilde
      \epsilon_3 \widetilde \epsilon_2  
      & c_0 |\vec\Lambda|^2 + c_1 \widetilde \epsilon_3 \widetilde \epsilon_3 
     \end{array} \right) + \cdots
\end{eqnarray}
where the $\widetilde \epsilon_i$ were defined before in
Eq.~(\ref{eq:TildeEpsilon}). Coefficients $c_0$ and $c_1$ contain
couplings and supersymmetric masses. Since they cancel in the final
expression for the angle their exact definition is not necessary in
the following. Dots stand for other terms which we will assume to be
less important in the following.  This matrix can be diagonalized
approximately taking in account that
\begin{eqnarray}
  \label{eq:3}
  x\equiv\frac{c_1 |\vec{\widetilde{\epsilon}}|^2}{c_0 |\vec \Lambda|^2 }\ll 1
\end{eqnarray}
Then
\begin{eqnarray}
  \label{eq:5}
  \widetilde m_\nu= c_0 |\vec\Lambda|^2 
      \left( \begin{array}{ccc} \displaystyle
      x\, \frac{\widetilde \epsilon_1 \widetilde \epsilon_1 }{
      |\vec{\widetilde \epsilon|^2} } 
      &\displaystyle x\, \frac{\widetilde \epsilon_1 \widetilde
      \epsilon_2 }{|\vec{\widetilde \epsilon|^2}} 
      & \displaystyle x\, \frac{\widetilde \epsilon_1 \widetilde
      \epsilon_3 }{|\vec{\widetilde \epsilon|^2}} \\[+4mm]
      \displaystyle
      x\, \frac{\widetilde \epsilon_2 \widetilde \epsilon_1 }{
      |\vec{\widetilde \epsilon|^2 }}
      &\displaystyle x\, \frac{\widetilde \epsilon_2 \widetilde
      \epsilon_2 }{|\vec{\widetilde \epsilon|^2}}
      & \displaystyle x\, \frac{\widetilde \epsilon_2 \widetilde
      \epsilon_3 }{|\vec{\widetilde \epsilon|^2}} \\[+4mm] 
      \displaystyle
      x\, \frac{\widetilde \epsilon_3 \widetilde \epsilon_1 }{
        |\vec{\widetilde \epsilon|^2 }}
      &\displaystyle x\, \frac{\widetilde \epsilon_3 \widetilde
      \epsilon_2 }{|\vec{\widetilde \epsilon|^2}}
      & \displaystyle 1 + x\, \frac{\widetilde \epsilon_3 \widetilde
      \epsilon_3 }{|\vec{\widetilde \epsilon|^2}}  
     \end{array} \right)
\end{eqnarray}

\begin{figure}[htbp]
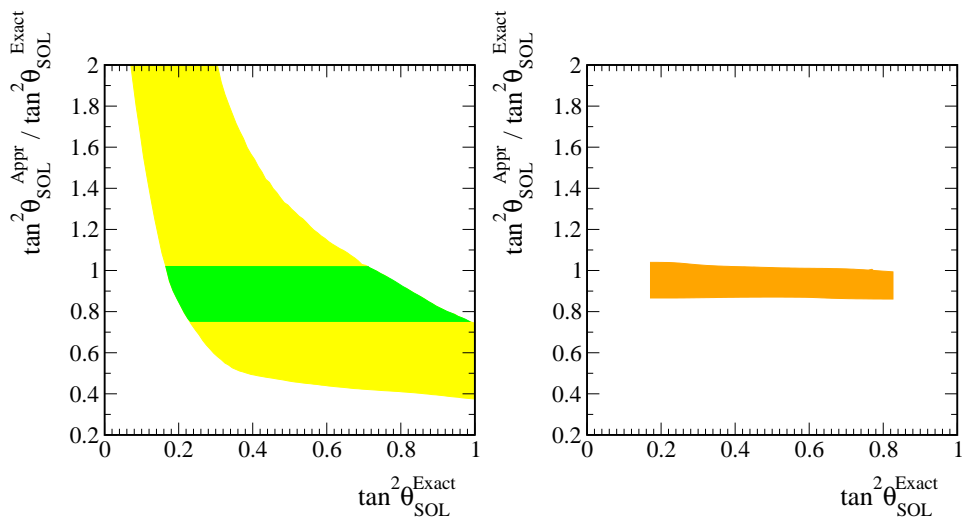
 \centering
    \includegraphics[width=0.4\linewidth]{Nfnd0Angle-v3.eps}
    \includegraphics[width=0.4\linewidth]{Nfnd1Angle-v3.eps}
\caption[] {(${ \tan^2\theta_\Sol }^{Appr}/
  {\tan^2\theta_\Sol}^{exact}$) versus ${\tan^2\theta_\Sol}^{exact}$.
  On the left panel the darker region contains over 90\% of the points
  in our sample. In the right panel the points in the region shown
  satisfy the cut $\sin(2\theta_{\tilde b}) \Delta
  B_0^{\tilde\tau_2\tilde\tau_1}> 0.02$ .}
\label{fig:SolarAngle}
\end{figure}

The rotation matrix that diagonalizes $\widetilde m_\nu$ in
Eq.~(\ref{eq:5}) can be written as
\begin{eqnarray}
  \label{eq:9}
  \widetilde V_\nu^T \widetilde m_\nu \widetilde V_\nu
  =\hbox{diag}(m_1,m_2,m_3) 
\end{eqnarray}
where
\begin{eqnarray}
  \label{eq:10}
   \widetilde V_\nu^T= \left( \begin{array}{ccc}
       e_{1,1} & e_{1,2}  & e_{1,3} \\
       e_{2,1} & e_{2,2}  & e_{2,3} \\
       e_{3,1} & e_{3,2}  & e_{3,3} 
       \end{array}\right)
\end{eqnarray}
The lepton mixing matrix is then given by
\begin{eqnarray}
  \label{eq:11}
  U=\left(  V_\nu^T  \widetilde V_\nu^T\right)^T
\end{eqnarray}
The expression for the solar mixing angle can be obtained from:
\begin{eqnarray}
  \label{eq:sol1}
  \tan^2\theta_\Sol  = \frac{U_{e2}^2}{U_{e1}^2}
\end{eqnarray}
From the above equations we obtain the very simple expression for the
solar mixing angle,
\begin{eqnarray}
  \label{eq:12}
   \tan^2\theta_\Sol  = \frac{\widetilde  \epsilon_1^2}{\widetilde
  \epsilon_2^2} 
\end{eqnarray}
This formula is a very good approximation if the one--loop matrix has
the structure $\epsilon_i \times \epsilon_j$, as is the case of the
bottom-sbottom loop if $m_{\nu_3} \gg m_{\nu_2}$, as illustrated in
Fig.~\ref{fig:SolarAngle}.

In the left panel we show a calculation comparing for all points the
approximate to the exact solar angle in the set with neutralino LSP,
while the right panel shows a subset of points with the cut
$\sin(2\theta_{\tilde b}) \Delta B_0^{\tilde\tau_2\tilde\tau_1}>0.02$.
Note that this cut is designed so as to favor points in which there is
a sizeable bottom-sbottom loop contribution to the full one-loop
neutrino mass. One sees from the right panel that for this case the
true solar angle is well approximated by our analytical formula.
Note finally that eq. (\ref{eq:12}) will fail completely, if 
$\Lambda_{\mu} \equiv \Lambda_{\tau}$ and $\epsilon_{\mu} \equiv 
\epsilon_{\tau}$, since then ${\widetilde \epsilon_2^2} =0$, see 
Eq. (\ref{eq:TildeEpsilon}). This is the origin of the ``sign 
condition'' discussed in \cite{hirsch:2000ef}. 

\section{Testing neutrino properties at high energy accelerators}
\label{sec:test-neutr-prop}

Since R-parity is broken in our model, the lightest supersymmetric 
particle is unstable and decays. This leads to the exciting possiblity 
to test the bilinear model at future colliders, such as the LHC or 
a possible Linear Collider.

The principle idea of such a collider test 
\cite{porod:2000hv,hirsch:2002ys,Hirsch:2003fe} is easily understood: 
Bilinear R-parity breaking leads to mixing between particles and 
sparticles with the same quantum numbers, as discussed above extensively 
for the case of neutrinos/neutralinos. This mixing, however, is not 
arbitrarily different for each particle/sparticle species. In fact, 
the bilinear model has just six new parameters, which we choose to be 
$\epsilon_i$ and $\Lambda_i$, compared to the MSSM. Essentially five 
of these six can be fixed from neutrino physics. 

Thus, if the MSSM parameters were known, all mixing effects could be 
calculated and thus all decay properties of the LSP would be fixed - 
apart from the effects of the last unknown parameter. In reality, however, 
the MSSM soft SUSY breaking parameters are completely {\em unknown}. The 
approach taken in \cite{porod:2000hv,hirsch:2002ys,Hirsch:2003fe} therefore 
is to calculate ratios of branching ratios of different decays. By 
taking ratios one essentially scales out the unknown MSSM parameters 
approximately and obtains observables which are proportional to either 
$\Lambda_i/\Lambda_j$ or $\epsilon_i/\epsilon_j$ (or some weird combination 
thereof). Which ratio one measures depends of course on the final 
state and the LSP under consideration. Since ratios of $\Lambda_i$'s (or 
$\epsilon_i$'s) are correlated with the neutrino angles, as discussed 
above, fixing neutrino angles from experimental data therefore gives 
definite predictions for some ratios of branching ratios.

\begin{figure}[htbp] \centering
    \includegraphics[width=0.4\linewidth]{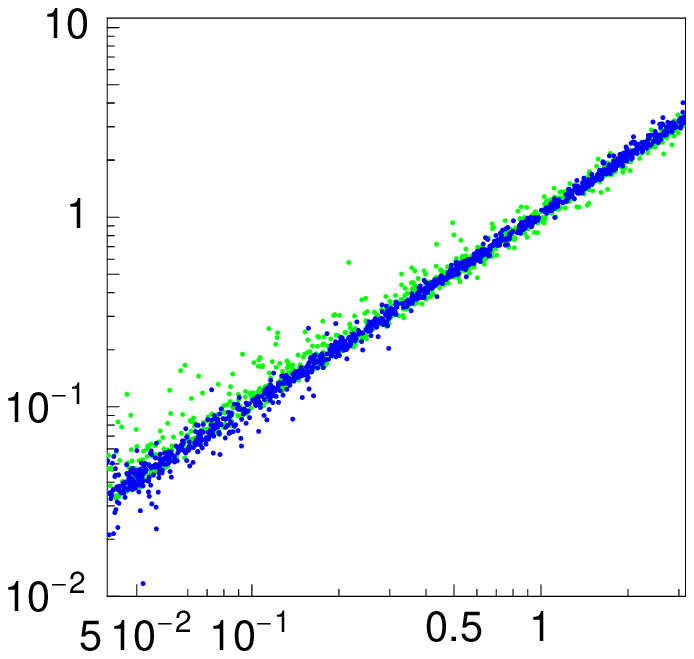}
    \includegraphics[width=0.4\linewidth]{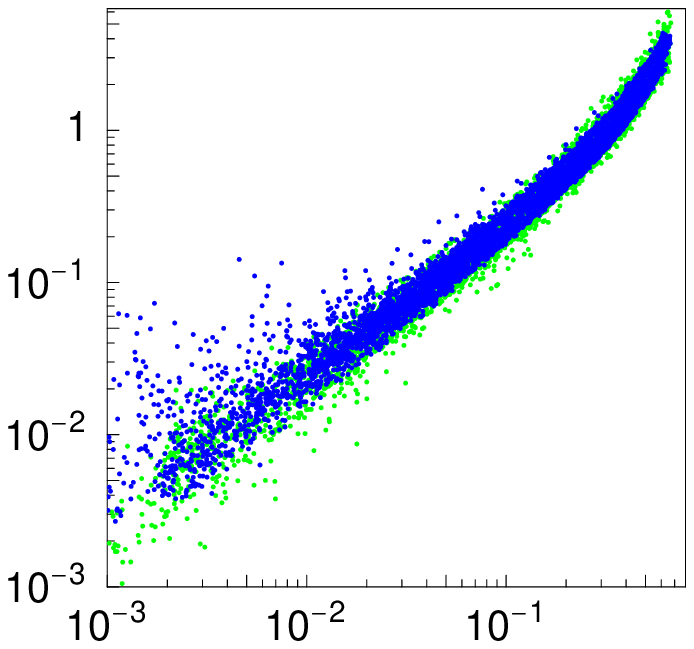}
\caption[] {Ratio of branching ratios of neutralino LSP decays. 
To the left: Br($\mu q q'$)/Br($\tau q q'$) as a function of 
$\tan^2(\theta_{atm})$. To the right: Br($e q q'$)/Br($\tau q q'$) as 
a function of $U_{e3}^2$. Whether bilinear R-parity breaking SUSY is 
responsible for atmospheric neutrino oscillations can be checked by 
such a measurement. Note that the spread of the points is 
entirely due to the unknown MSSM parameters. Even a moderately 
accurate input of MSSM parameters will lead to much sharper predictions 
for these decays.}
\label{fig:NtrlDec}
\end{figure}

One example is shown in Fig. \ref{fig:NtrlDec}, where ratio of branching 
ratios of neutralino LSP decays are plotted. Note that 
Br($\mu q q'$)/Br($\tau q q'$) is directly proportional to 
$\tan^2(\theta_{atm})$, i.e. should be near $\sim 1$ according to 
current neutrino data. The spread in the points is due to the unknown 
MSSM parameters. Of course, once SUSY is discovered these unknowns 
will be measured allowing for much sharper tests of the model than 
indicated in Fig. \ref{fig:NtrlDec}.

\begin{figure}[htbp] \centering
    \includegraphics[width=0.4\linewidth]{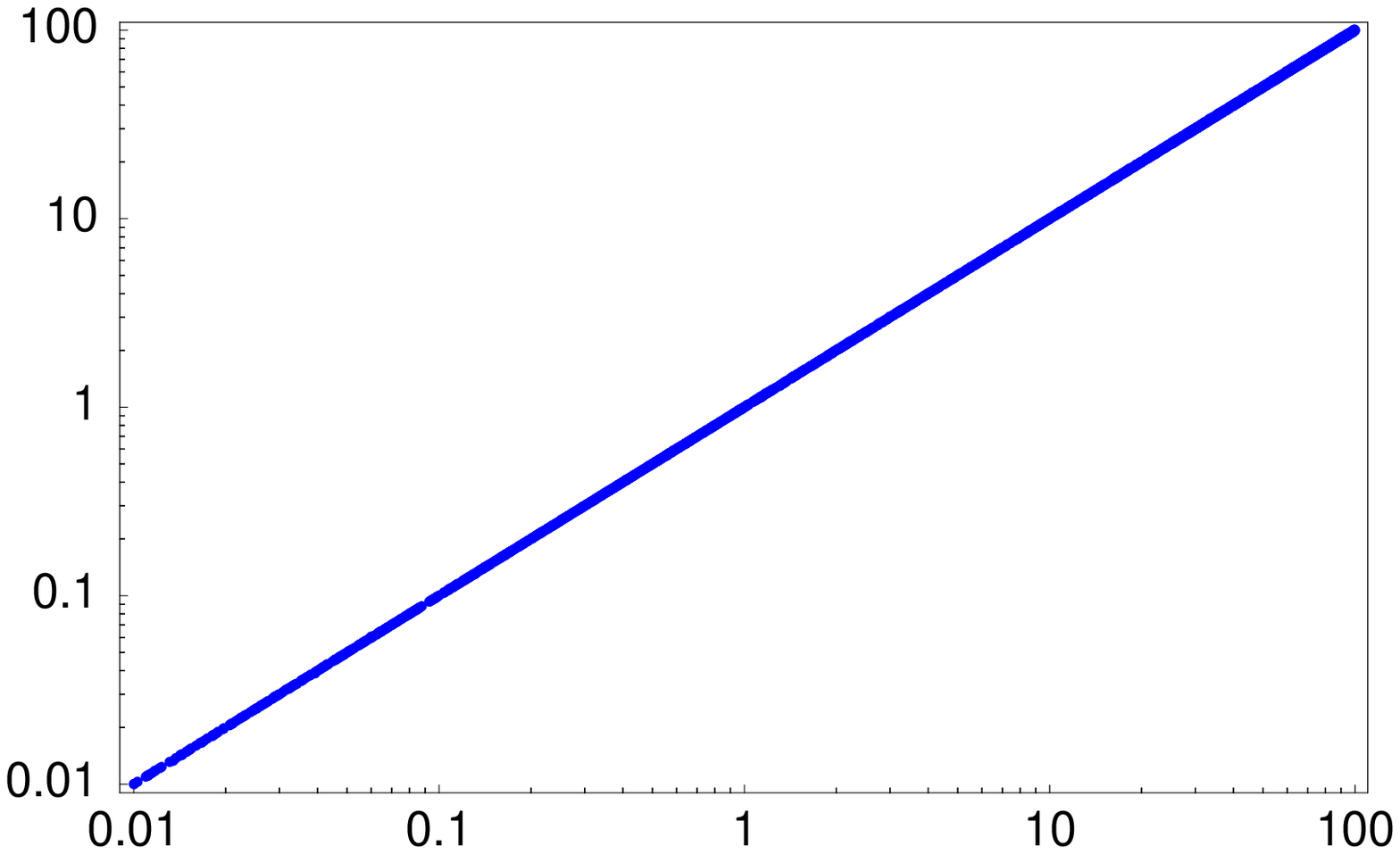}
    \includegraphics[width=0.4\linewidth]{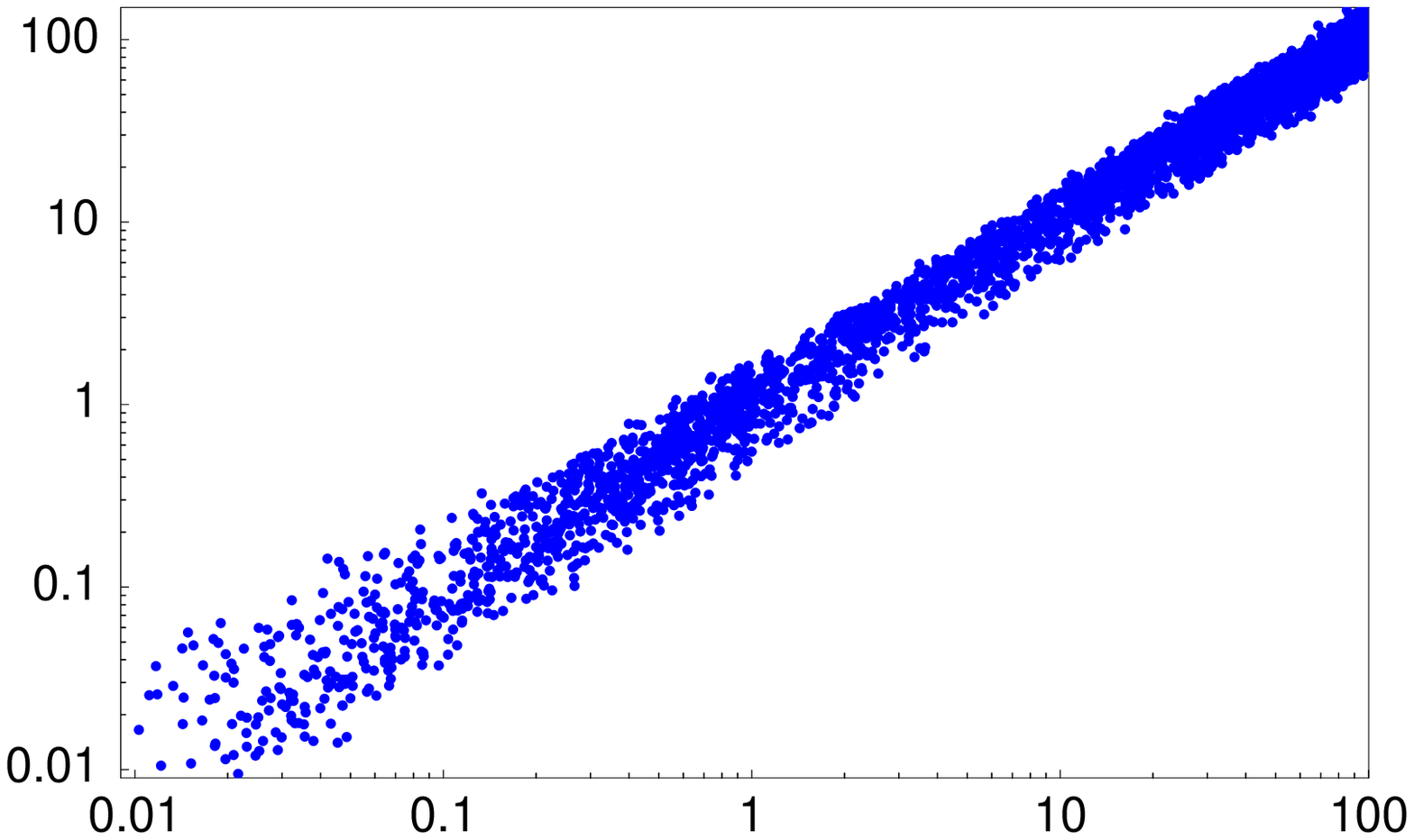}
\caption[] {Ratio of branching ratios of scalar tau LSP decays. 
To the left: Br(${\tilde \tau}_1 \to  \sum \nu e$)
/Br(${\tilde \tau}_1 \to  \sum \nu \mu$) as a function of 
$(\epsilon_1/\epsilon_2)^2$. To the right: as a function of 
$\tan^2(\theta_{\odot})$. Whether bilinear R-parity breaking SUSY is 
responsible for solar neutrino oscillations can be checked by 
such a measurement if scalar taus are the LSP. }
\label{fig:StauDec}
\end{figure}

A second example is shown in Fig. \ref{fig:StauDec}, where we show 
Br(${\tilde\tau}_1 \to  \sum \nu e$)/
Br(${\tilde\tau}_1 \to  \sum \nu \mu$) as a function of 
$(\epsilon_1/\epsilon_2)^2$ (left) and as a function of 
$\tan^2(\theta_{\odot})$ 
(right). Obviously this ratio is strongly correlated with the solar 
angle and thus, if scalar taus turn out to be the LSP, such a 
measurement would provide an excellent check of the bilinear model 
as the origin of the solar neutrino mass scale.

With the LSP unstable, in principle any sparticle can be the LSP. 
In \cite{Hirsch:2003fe} the remaining candidates have been discussed: 
Charginos, scalar quarks, gluinos and scalar neutrinos. 
The main conclusion of \cite{Hirsch:2003fe} is that whichever SUSY particle 
is the LSP, measurements of branching ratios at future accelerators 
will provide a definite test of bilinear R-parity breaking as 
the model of neutrino mass. We just mention that chargino LSPs 
would be more sensitive to atmospheric neutrino physics (as are neutralinos) 
while the other LSP candidates mentioned above show more dependence 
on the solar neutrino angle.

\section{Discussion and conclusions}
\label{sec:disc-concl}

We have presented a brief review of the idea that supersymmetry with
explicit bilinear breaking of R-parity is the origin of neutrino masses 
and lepton mixing.
The bilinear R-parity breaking
(BRpV) model is the simplest extension of the minimal supersymmetric
standard model (MSSM) which includes lepton number violation.
We have seen how it leads to a successful phenomenological model for
neutrino oscillations, in accordance to present neutrino data. The
pattern of neutrino masses is hierarchical, with the atmospheric mass
scale arising at the tree level whereas the solar scale is induced
from calculable loop corrections.
We saw how, in contrast to seesaw models, the BRpV model can be probed
at future collider experiments, like the LHC or the NLC. Indeed we
have discussed how, irrespective of the supersymmetric particle
which is the lightest, its decay pattern will be directly related with
the lepton mixing angles determined in low energy neutrino
experiments.

\section*{Acknowledgements}  

This work was supported by Spanish grant BFM2002-00345, by the
European Commission RTN grant HPRN-CT-2000-00148 and the ESF
\emph{Neutrino Astrophysics Network}.  M. H.  is supported by a
Ramon y Cajal contract.

\end{document}